\documentclass[doublecol]{epl2}
\newcommand{\M}{{\cal M}}
\newcommand{\rhoL}{{\rho_\Lambda}}

\renewcommand{\d}{{\rm d}}
\newcommand{\mub}{\bar\mu}
\newcommand{\bi}{{\rm bi}}

\newcommand{\half}{{\frac{1}{2}}}

\newcommand{\myskip}[1]{}    
       
\newcommand{\p}{\partial}

\newcommand{\BEQ}{\begin{eqnarray}}   
\newcommand{\EEQ}{\end{eqnarray}}   
\newcommand{\BEA}{\begin{eqnarray}}   
\newcommand{\EEA}{\end{eqnarray}}   
\newcommand{\nn}{\nonumber }   
\renewcommand{\d}{{\rm d}}

\newcommand{\mn}{{\mu\nu}}   

\newcommand{\lambdaB}{{\bar\lambda}}  

\begin{document}

\title{Supermassive Black Holes as Giant Bose-Einstein Condensates}

\shorttitle{Bose-Einstein Condensed Supermassive Black Hole}
\author{Theo M. Nieuwenhuizen}
\shortauthor{Th.M. Nieuwenhuizen} %, resubmitted on, 2008}

\institute{Institute for Theoretical Physics,  University of Amsterdam,   
Valckenierstraat 65, 1018 XE Amsterdam, The Netherlands}


\abstract{
The Schwarzschild metric has a divergent energy density at the horizon, 
which motivates a new approach to black holes.
If matter is spread uniformly throughout the interior of a supermassive 
black hole, with mass $M\sim M_\star= 2.34\,10^8M_\odot$,
 it may arise from a Bose-Einstein condensate of densely packed H-atoms. 
Within the Relativistic Theory of Gravitation with a positive cosmological 
constant, a bosonic quantum field is coupled to the curvature scalar.
In the Bose-Einstein condensed groundstate an exact, selfconsistent solution for 
the metric is presented.
It is regular with a specific shape at the origin. 
The redshift at the horizon is finite but large, $z\sim 10^{14}$$M_\star/M$. 
The binding energy remains as an additional parameter to characterize the BH; 
alternatively, the mass observed at infinity can be any fraction of the rest mass
of its constituents.
}

\maketitle
%\email{t.m.nieuwenhuizen@uva.nl}
 
On the basis of the Schwarzschild, Kerr and Kerr-Newman metrics, it generally believed 
that black holes (BHs) are singular objects with all matter localized in the center
or, if rotating, on an infinitely thin ring.
Recent approaches challenge this unintuitive assumption and consider matter just spread  
throughout the interior.~\cite{Laughlin,Dymnikova,MazurMottola} 
Here we shall follow this line of research.
To start, let us just look at some orders of magnitude.
For solar mass neutron stars it is known that the density is about the nuclear density.
Solar mass black holes are about $10$ times smaller, thus $1000$ times more compact.
Clearly, this begs for a Quantum Chromodynamics description in curved space. 

We shall focus on the other extreme, superheavy BHs. They occur
in the center of each galaxy and weigh about 
$M_{\rm BH}=0.0012\,M_{\rm bulge}$. \cite{BHmassBulgeMass}
Let us assume that they consist of hydrogen atoms
and that mass and particle number are related as $M\equiv \nu Nm_{\rm H}$ with some $\nu\le 1$. 
If we neglect rotation, the radius is $R=GM/c^2$ (see below).
We may compare the BH density $3N/4\pi R^3$ with
the one of densely packed, non-overlapping  H-atoms, that is,
with the Bohr density $n_{\rm B}\equiv 3/4\pi a_0^{3}$, with $a_0=0.529\,\AA$ the Bohr radius. 
This yields a mass $M_\star=c^3(a_0^3/G^3m_{\rm H})^{1/2}$ (we take $\nu=1$ here), which lies 
in the range of observed supermassive black holes, 
$M_{\star}= 4.66 \,10^{38} {\rm  kg}=2.34  \, 10^8\,M_\odot$.

Next comes the question how matter can withstand the enormous pressure
normally associated with such high densities. It was proposed originally by
Sacharov that the vacuum equation of state $p=-\rho$ could describe 
matter at superhigh densities.~\cite{Sacharov}
Laughlin and coworkers assume that matter near the horizon could be in 
its Bose-Einstein condensed (BEC) phase, modeled by the 
vacuum equation of state. ~\cite{Laughlin} 
Dymnikova considers BHs obeying it in the interior,
which, however, have one or two horizons.~\cite{Dymnikova} 
Mazur and Mottola take the BEC idea over to the interior, and
investigate a ``gravo-star'', of which the interior obeys the vacuum equation of state, 
and which is surrounded by a thin shell of normal matter having the stiff equation 
of state $p=+\rho$. This solution is regular everywhere.~\cite{MazurMottola}

We shall demonstrate that a supermassive BH can exist as a self-gravitating hydrogen cloud,
in a Bose-Einstein condensed phase. 
We study the problem in a series of improved starting points:
assume a stiff equation of state,  self-consistently solve 
a quantum field coupled to the curvature scalar,  
first for a uniform groundstate wavefunction and next for a space-dependent one.
Hereto we have to employ the Relativistic Theory of Gravitation
(RTG), which reproduces all weak gravitational effects in the solar system
~\cite{LogunovBook,LogInflaton} as well as the $\Lambda$CDM cosmology \cite{NEPL}.

We consider a static metric with spherical symmetry, 

\BEQ \label{sphersymmetric}
\d s^2=U(r)c^2\d t^2-V(r)\d r^2-W^2(r)
(\d\theta^2+\sin^2\!\theta\d\phi^2).\EEQ 
The gravitational energy density arises from the Landau-Lifshitz pseudo-tensor,
\cite{LandauLifshitz} generalized to become a tensor in Minkowski space.
~\cite{BabakGrishchuk,NEPL}
For (\ref{sphersymmetric}) it takes the form 
\BEQ \label{t00r=}
t^{00}&=&\frac{c^4W^2}{8\pi Gr^6}
\left(-\frac{r^2V'WW'}{V}+r^3V' -5r^2W'{}^2\right.\nn\\
&+&
\left.\frac{2r^3VW'}{W}+8rWW'-2r^2V-3W^2\right).
\EEQ
Let us start with the General Theory of Relativity (GTR).
The Schwarzschild metric reads in the harmonic gauge
\BEQ \label{Schwarzharm}
U_S=\frac{1}{V_S}=1-\frac{2M}{W_S}=\frac{r-M}{r+M},\quad W_S=r+M.
\EEQ
(We put $G=c=\hbar=1$.) 
It is singular at the horizon $r_h=M$
and involves the gravitational energy density 
\BEQ     \label{ThetaSS}
t^{00}=\frac{1}{4\pi r^2}\frac{\d}{\d r}\frac{M(r+M)^3(2r+M)}{2r^3(r-M)}.
\EEQ
Its quadratic divergence at $r_h$ presents a hitherto overlooked peculiarity, 
that induces a negative infinite contribution to the total energy. 
For this reason, we shall switch to RTG with matter not located at the singularity $r=0$,
but just spread out within the horizon.

\subsection{Quantum field theory of Bose-Einstein condensed black holes}
Let our H-atoms be described by a bosonic field 
\BEQ 
\hat \psi({\bf r},t)=\sum_i \hat a_i\psi_i({\bf r})e^{-iE_it},
\nn
\EEQ
where $i=\{n,\ell,m\}$, 
$[\hat a_i, \hat a_j^\dagger]=\delta_{ij}$  and eigenfunctions factor as
$\psi_i({\bf r})=\phi_n(r)$$Y_{\ell m}(\theta,\phi)$.
The rotating wave approximation then leads to the Lagrangian
~\cite{StringariPitaevski}

\BEQ \label{Lagm=}
L_{\rm mat}=g^\mn\p_\mu\hat \psi^\dagger\p_\nu\hat\psi
-(m^2+\xi R)\hat\psi^\dagger\hat\psi
-\frac{\lambda}{4}\hat\psi^\dagger{}^2\hat\psi^2. 
\EEQ 
For a field in curved space the renormalization group 
generates a coupling to the Ricci curvature scalar $R$.~\cite{Birell}
Its strength $\xi$ is for now a phenomenological parameter.
The dimensionless coupling $\lambda=8m^2c g/\hbar^3$ with 
$g=4\pi\hbar^2 a_s/m$  models the two particle interaction by the scattering 
length. For hydrogen in flat space one has~\cite{Stoof}

\BEQ a_s=0.32\,a_0\quad\textrm {singlet state},\qquad
 a_s=1.34a_0 \quad \textrm {triplet}. \nn \EEQ
We shall continue with the singlet value $\lambda=0.81\,10^7$.

With $\Psi_0=({2E_0N_0})^{1/2}$$\psi_0e^{-iE_0t}$ for 
$N_0$ groundstate atoms, the relativistic Gross-Pitaevskii equation reads 

\BEQ \label{GPeq}
\left(\frac{1}{\sqrt{-g}}\p_\mu \sqrt{-g}g^\mn\p_\nu+m^2+\xi R+
\frac{\lambda|\Psi_0^2|}{4E_0}\right)\Psi_0 =0. 
\EEQ

A homogeneous ground state, $\Psi_0(r,t)=\Psi_0e^{-iE_0t}$ 
occurs when
\BEQ \label{GScond}
-\frac{E_0^2}{U}+m^2+\xi R+\frac{\lambda}{4E_0}|\Psi_0|^2=0.
\EEQ
We focus on RTG, which describes gravitation as a field in Minkowski space
 ~\cite{LogunovBook,LogInflaton} and posesses the same gravitational energy 
momentum tensor and thus also the gravitational energy density Eq. (\ref{t00r=}).~\cite{NEPL}
It extends the Hilbert-Einstein action with 
the cosmological term and a bimetric coupling between the
Minkowski ($\gamma$) and Riemann ($g$) metrics, 

\BEQ \label{Lagtot=}
L&=&-\frac{R}{16\pi}-\rho_\Lambda+\half\rho_\bi \gamma_\mn g^\mn+L_{\rm mat}.
\EEQ
(For $\rho_\bi=0$ it is just a field theory for GTR.) One has
\BEQ \label{rhotot=}
\rho_{\rm tot}&=&\rho+\rhoL
+\frac{\rho_\bi}{2U}-\frac{\rho_\bi}{2V}-\frac{\rho_\bi r^2}{W^2},\nn\\
p_{i}^{\rm tot}&=&
p_{i}-\rhoL+\frac{\rho_\bi}{2U}-\frac{\rho_\bi}{2V}+\frac{\rho_\bi r^2}{W^2}.
\EEQ
with $i=r,\theta,\phi$.
The value $\rho_\bi=\rho_\Lambda$ is imposed to have a Minkowski metric in the absence of matter.
One may fix them to the observed positive cosmological constant~\cite{NEPL}. 
However, historically the opposite choice $\rho_\Lambda<0$ was considered
and the cosmological data were described by an additional inflaton field.~\cite{LogInflaton}
So the sign of $\rho_\bi$ is not known yet; We show
 that solving a realistic black hole settles this issue.
The new point of RTG is that $g_{00}=U$ can be very small.
Despite the smallness of $\rho_\bi$, 
the $\rho_\bi/U$ terms become relevant near the horizon~\cite{LogunovBook,NEPL}, bringing

\BEQ 
R=-8\pi T_{\rm tot}=8\pi(-\rho+p_r+p_\theta+p_\phi+\frac{\rho_\bi}{U}). 
\EEQ

To start, let us consider the stiff equation of state 
\BEQ \label{rhopU} 
\rho=\half \rho_c(\frac{U_c}{U}+1),\qquad
p_i=p\equiv\half \rho_c(\frac{U_c}{U}-1),
\EEQ
where $\rho_c=3/32\pi M^2$. For $U_c=0$ it is the vacuum equation of state $p=-\rho=$const. 
One gets
\BEQ R=8\pi\frac{\rho_cU_c+\rho_\bi}{U}-16\pi\rho_c.\EEQ
Eq. (\ref{GScond}) has a solution due to the $\xi R$-term. 
The constant and $1/U$ terms term imply, 
\BEQ \label{xi=}
 \xi=\xi_0(1+\frac{\lambda|\Psi_0|^2}{4E_0m^2}),\qquad
 E_0^2=8\pi\xi(\rho_cU_c+\rho_\bi),
\EEQ
respectively. The dimensionless parameter
\BEQ
\xi_0=\frac{2}{3}m^2M^2= 1.80 
\, 10^{54}\left(\frac{M}{M_\star}\right)^2, 
\EEQ
appears to be large, but since $R\sim1/M^2$ 
the combination $\xi R$ is just of order $m^2$, making its effect of order unity.
% From now on, we take $\xi$ as a phenomenological parameter.
The metric can be solved as below; we omit details.

\subsection{Self-consistent field theory}
 Rather than imposing an equation of state, 
the material energy-momentum tensor $T^\mn_m$ should be derived from first principles, i.e. 
from the quantum field theory for the H-atoms.  
Its energy density reads, if we exclude the effect of the $\xi R$-term,
\BEQ \label{rhopgen=}
\rho_m&=&\frac{\langle\p_t\psi^\dagger\p_t\psi\rangle}{U}
+\frac{\langle\p_r\psi^\dagger\p_r\psi\rangle}{V}
+\frac{\langle\p_\theta\psi^\dagger\p_\theta\psi\rangle}{W^2}\nn\\
&+&
\frac{\langle\p_\phi\psi^\dagger\p_\phi\psi\rangle}{W^2\sin^2\theta}
+m^2\langle\psi^\dagger\psi\rangle
+\frac{\lambda}{4}\langle\psi^\dagger{}^2\psi^2\rangle.
\EEQ
The pressures $(p^m_{r},p^m_{\theta},p^m_{\phi})$ have this shape 
with signature $(++----)$, $(+-+---)$, and $(+--+--)$, respectively.
Spherical symmetry will imply that $p^m_\theta=p^m_\phi\equiv p^m_\perp$.
For a uniform groundstate $p_m$ is isotropic,
\BEQ \label{rhopground=}
(\rho_m,p_m)=\half(\frac{E_0}{U}\pm\frac{m^2}{E_0})|\Psi_0^2|
\pm\frac{\lambda|\Psi_0^4|}{16E_0^2}.
\EEQ 
They consist of a vacuum part $p=-\rho=$ const. 
and a stiff part $p=+\rho\sim1/U$, the types studied in 
~\cite{MazurMottola} and above. 
In the non-relativistic ($E_0=m$) and flat space ($U=1$) limit, 
they reduce for $\lambda=0$ to $\rho_m=mc^2|\Psi_0^2|$ and $p_m=0$.

Because of the $\xi R$-term in (\ref{Lagm=}), 
the Einstein equations embody a direct backreaction of matter on curvature,
$G^\mn=8\pi(T^\mn_m+T^\mn_\Lambda+T^\mn_\bi)-16\pi\xi\langle\psi^\dagger\psi\rangle G^\mn$.
To connect to the standard notation, $G^\mn=8\pi T^\mn_{\rm tot}$,
we define $T^\mn$  by
\BEQ \label{Einsteff=} 
T^\mn_{\rm tot}=\frac{T_m^\mn+T^\mn_\Lambda+T^\mn_\bi}{1+B}
\equiv  T^\mn+T^\mn_\Lambda+T^\mn_{\bi},
\EEQ
with direct backreaction strength of matter on the metric

\BEQ \label{Bdef}
B=16\pi\xi\langle\psi^\dagger\psi\rangle=\frac{8\pi \xi |\Psi_0^2|}{E_0}.
\EEQ
For $\lambda=0$ the curvature scalar follows from (\ref{rhopground=}) as

\BEQ R=\frac{8\pi}{1+B} 
 \left(\frac{E_0|\Psi_0^2|+\rho_\bi}{U}-\frac{2m^2}{E_0}|\Psi_0^2|\right). 
\EEQ
Solving Eq. (\ref{GScond}), we find two relations and a consequence,
\BEQ \label{Psiconds}
B=1,
\quad E_0^2=8\pi\xi\rho_\bi,\quad
|\Psi_0^2|=\frac{E_0}{8\pi\xi}=\frac{\rho_\bi}{E_0}.
\EEQ
(The GTR situation, reached by taking $\rho_\bi\to 0$ first, would 
not allow a meaningful solution.) 
The first identity expresses a $100\%$ direct backreaction
of matter on the metric. 
This motivates to introduce the parameters ~\cite{NEPL}
\BEQ \mu=\sqrt{16\pi\rho_\bi}=\sqrt{2\Lambda},\qquad
\mub=\mu M =7.90\,10^{-15}\frac{M}{M_\star}. 
\EEQ
Instead of searching a finite $U$, as for boson stars, ~\cite{bosonstar} 
we assume a very small $U$ with $U(0)=0$, coded by 
%a parameter 
$\upsilon$, 
%(``upsilon''),
\BEQ \label{U=}
 U=\frac{1}{2}\mu^2\upsilon^2W^2
\EEQ
In terms of the mass function $\M(r)$, defined by 
\BEQ \label{VMdef} V=\frac{W'{}^2}{1-2\M/W},\EEQ
the $00$ and $11$ Einstein equations take the form
\BEQ \label{M'=}
\M'=4\pi W' W^2\rho_{\rm tot},\quad \!\!
\frac{W-2\M}{2UW^2}\frac{U'}{W'}-\frac{\M}{W^3}=
4\pi \,p^r_{\rm tot}.
\EEQ
The Ansatz (\ref{U=}) solves them and yields, due to (\ref{Psiconds}),
\BEQ \label{Einout}
\upsilon=1,\qquad \M=\frac{W}{4}+\frac{W^3}{16M^2}\frac{2m^2M^2}{3\xi}.
\EEQ
For the Schwarzschild black hole the horizon occurs when $\M=M$ for $W=2M$.
Concerning the outside metric, we will be close to that situation.
This implies again that a mass $M$ corresponds to $\xi=2m^2M^2/3$. 

Let us introduce the `Riemann' variables $x$ and $y$ by
\BEQ 
x=\frac{W}{2M},\qquad y=\sqrt{1-x^2},\qquad  
\EEQ
so that $U=2\mub^2x^2$. 
The $\rho_\bi$ terms in (\ref{Lagtot=}) violate general coordinate invariance 
and impose the harmonic gauge, 

\BEQ \label{harm}
\frac{U'}{U}-\frac{V'}{V}+4\frac{W'}{W}=\frac{4rV}{W^2}.
\EEQ
With (\ref{U=}), (\ref{VMdef}) and $\M=\half M(x+x^3)$ from (\ref{Einout}), 
it brings
\BEQ 
\frac{2x'}{x}-\frac{2x''}{x'}-\frac{2xx'}{1-x^2}
+\frac{4x'}{x}=\frac{8rx'{}^2}{x^2(1-x^2)}. \nn
\EEQ
Going to the inverse function $r(x)$ makes it linear,
\BEQ x^2(1-x^2)r''+x(3-4x^2)r'=4r. 
\EEQ
The solution is then remarkably simple,

\BEQ\label{rx=}
r=r_1(1+\frac{y}{\sqrt{5}}) x^{\sqrt{5}-1}(1+y)^{-\sqrt{5}},
\qquad \label{y=}
\EEQ
(The second independent solution with $\sqrt{5}\to-\sqrt{5}$ is singular.)
This determines the metric function $V$,
\BEQ 
%\label{W'x=} % W'&=&\frac{M\sqrt{5}}{2r_1}x^{2-\sqrt{5}}y(1+y)^{\sqrt{5}},\nn\\
\qquad \label{Vx=}
V&=&\frac{2W'{}^2}{y^2}=\frac{5M^2}{2r_1^2}x^{4-2\sqrt{5}}(1+y)^{2\sqrt{5}}.
\EEQ

Putting these results together, it now follows that
\BEQ 
 \rho=\frac{3}{64\pi M^2} ,\quad  p=-\frac{3}{64\pi M^2},
\EEQ 
as the $1/U$ terms cancel due to the relation $E_0^2=8\pi\xi\rho_\bi$.
So, after all, we reproduce the vacuum equation of state.

To normalize $|\Psi_0|$, we need
the $3d$ volume element in the future time direction,
$\d \Sigma^\mu= \d r\d\theta\d\phi n^\mu \sqrt{-g_3}\equiv \delta^\mu_0\d {\cal V}$,
set by the timelike unit vector $n^\mu=\delta^\mu_0/\sqrt{U}$
and $g_3=-VW^4\sin^2\theta$. This results in
$\d {\cal V}=\d r \d\Omega \sqrt{V/U\,}\,W^2$.

The general inner product ~\cite{Birell} $(\psi_1,\psi_2)=-i\int \d \Sigma^\mu\times$ 
$(\psi_1\p_\mu\psi_2^\ast-\psi_2^\ast\p_\mu\psi_1)$ defines the orthonormality

\BEQ \label{innerprod}
(\psi_i,\psi_j)=(E_i+E_j)\int \d{\cal V} \,\psi_i\psi_j^\ast\equiv\delta_{ij}.
\EEQ
With $\d{\cal V}=\d y\d\Omega\,8M^3/(\mub\upsilon) $ it yields 

\BEQ \label{Psi02=}
|\Psi_0^2| =2E_0N_0|\psi_0^2|=\frac{N_0\upsilon\mub }{32\pi M^3},
\EEQ
having proper groundstate occupation,  $\int\d{\cal V}\,|\Psi_0|^2=N_0$.

In  Eq. (\ref{xi=}) the correction term is of order 
\BEQ 
\lambdaB \equiv \frac{\lambda}{32\pi m^2\xi} =
\frac{3\lambda}{64\pi\,m^4M^2}\approx 7.58 \,10^{-12} 
\frac{M_\star^2}{M^2}.
\EEQ
These corrections seem relevant for BH's with masses 
$M\sim 2.75\,10^{-6}M_\star\sim 645\,M_\odot$.
Much less below $M_\star$ the hydrogen atoms will get ionized, 
calling for fermionic fields for protons and electrons, 
which by a BCS pairing can again undergo a BEC transition. 
This BCS-BEC scenario is beyond the aim of the present paper.

%%%%%%%%%%%%%%%%%%%%%%%%%%%%%%%%%%%%%%%%%%%%%%%%%%%%%%%%%%%%%%%%%%

\subsection{The exterior}
At the horizon $r_h\approx r_1$, $y_h\ll 1$ one has
\BEQ \label{boundary}
U=2\mub^2,\quad V=\frac{5}{2},\quad W=2M,\quad W'=
\half\sqrt{5} \, y_h.
\EEQ
We have to connect this to the vacuum solution outside the BH.
Well away from matter, the harmonic constraint brings the Schwarzschild shape 
(\ref{Schwarzharm}), where $M\equiv\M(r_h)$ is the mass, essentially as observed 
at infinity. 
The values (\ref{boundary}) are far from Schwarzschild's,  
even when $r$ is near $M$ (e.g., $W_S'=1$).
The problem nevertheless appears to be consistent.
Near $r_h$ we need the deformation of the Schwarzschild metric 
which regularizes its singularity due to the bimetric coupling. 
~\cite{LogunovBook,LogInflaton}
An elegant scaling form for small $\mub$ was presented by us, \cite{NEPL}, 
\BEA\label{scaling}
r&=&M
\frac{1+\eta(e^\xi+\xi+\log\eta+2)}{1-\eta(e^\xi+\xi+\log\eta+2)},
\qquad
 U = \eta e^\xi, \\
V&=&\frac{e^ \xi}{\eta(1+e^\xi)^2},\qquad 
W=\frac{2M}{1- \eta e^\xi-\mub^2(\xi+w_0)}.\nn
\EEA
Here $\xi$ is the running variable and $\eta$ a small scale. 
For $\eta e^\xi={\cal O}(1)$ it coincides with the Schwarzschild solution.
Matching with the interior appears to be possible,
\BEQ  \label{vacuumvalues}
e^{\xi_h}=\sqrt{5}\mub,\quad 
\eta=\frac{2}{\sqrt{5}}\mub,\quad
W'=e^{\xi_h}+\frac{\mub^2}{\eta}=\frac{3}{2}\sqrt{5}\mub,
\EEQ
implying $y_h=3\mub$.
Taken together, the three regimes, interior, horizon and exterior, 
provide an exact solution of the problem.
At the origin it exhibits the singularities 
\BEQ U=\bar U_1r^{\gamma_\mu},\quad
V=\half\gamma_\mu^2\bar W_1^2r^{\gamma_\mu-2},\quad
W=\bar W_1r^{\half\gamma_\mu},\EEQ
where $\gamma_\mu=\half(\sqrt{5}+1)$ is the golden mean.
But if we take $W$ as the coordinate, we have in the interior the  shape
\BEQ
\d s^2=\half\mu^2W^2\d t^2-\frac{2\d W^2}{1-{W^2}/{4M^2}}-
W^2\d\Omega^2,
\EEQ
which is regular at its origin, with the term $2\d W^2$ coding the above singularities. 

We may rewrite the exterior solution by eliminating $\xi$,
\BEA\label{scalingU}
r&=&M
\frac{1+U+(2\mub/\sqrt{5})(\log U+2)}{1-U-(2\mub/\sqrt{5})(\log U+2)},
\nn\\ 
V &=&\frac{U}{(U+2\mub/\sqrt{5})^2},\qquad  
\\ 
W&=&\frac{2M}{1-U+2\mub^2 -\mub^2\log(U/2\mub^2)}. \nn
\EEA
This describes the free space region $r\ge M$, where $2\mub^2\le U\le 1+{\cal O}(\mub)$.
At the cosmic scale $r\sim1/\mu$ Newton's law picks up the Yukawa-type factor $\cos\mu r$,
due to the tachyonic nature of gravitation in RTG with $\rho_\bi>0$.~\cite{NEPL}

The interior shape can also be expressed with $U$ as running variable, 
where it lies in the range $(0, 2\mub^2)$.
Due to Eq. (\ref{y=}) it also holds that
\BEQ W=2Mx=\frac{1}{\mu}\sqrt{2 U},\qquad 
y=\sqrt{1-\frac{U}{2\mub^2}}.
\EEQ 
The locus and the metric function $V$ are given by (\ref{rx=}),
with $r_1\approx M$, and (\ref{Vx=}), respectively.

\subsection{Contributions to the energy}
With the weight $\d {\cal V}$ given below (\ref{innerprod}), the standard expression for the 
energy, $\int\d {\cal V}\rho$ scales as $1/\mub$ and even diverges logarithmically at $r=0$. 
However, in RTG the energy is determined by 
Eq. (\ref{t00r=}).
At the origin it diverges as $r^{\sqrt{5}-5}$, which is integrable. 
The gravitational energy inside the BH reads 
\BEQ
U_{\rm grav,\,int}=4\pi\int_0^M\d r\, r^2t^{00}=-842.898 %079
\,M.\nn
\EEQ
The total energy density reads 
$\Theta^{00}=t^{00}+{VW^4}\rho_{\rm tot}/{r^4}$.
We can calculate the material and the bimetric energy,
\BEQ
U_{\rm mat}&=&4\pi\int_0^M r^2\d r\frac{VW^4}{r^4}\,\frac{3}{64\pi M^2}= 169.431%6459
\,M, \nn \\
U_\bi&=&4\pi\int_0^M r^2\d r\frac{VW^4}{r^4}\,\frac{1}{64\pi M^2x^2}= 686.466%4330
\,M.\nn
\EEQ
Together they make up for 
\BEQ
U_{\rm interior}=U_{\rm grav,\,int}+ U_{\rm mat}+U_\bi=13 \,M. 
\EEQ
The energy density in the skin layer first has a large positive and then 
a large negative part, due to the term $r^3V'$. The integrated effect is
obtained easily since, in the formulation of the Einstein equations in Minkowski
space, the total energy density is a total derivative,~\cite{NEPL}
The region $r>M$ thus yields $U_{\rm exterior}=-12 M.$
Together with the interior it confirms the total energy $U=Mc^2$, expected from the
decay of the metric, $g_{00}=1-2GM/c^2r$.

%%%%%%%%%%%%%%%%%%%%%%%%%%%%%%%%%%%%%%%%%%%%%%%%%%%%%%%%%%%%%%%%%%%%%

\subsection{ Non-uniform groundstate} Till now we assumed that the groundstate 
wavefunction has a constant amplitude, and drops to zero at the horizon. 
Clearly, this cannot be exact. 
Taking $\Psi_0\to \Psi_0(r)$, we first take into account that in deriving the Einstein
equations, partial integrations are to be performed.  
This brings derivatives of $B\sim|\Psi_0|^2$, and induces an extra term $T^\mn_B$,
\BEQ \label{EinBeqs}
(1+B) G_\mn=8\pi(T_m^\mn+T_\Lambda^\mn+T_\bi^\mn+T_B^\mn).\EEQ
The elements of $(T_B)^\mu_\nu\equiv {\rm diag}(\rho_B,-p^B_r,-p^B_\perp,-p^B_\perp)$ are
\BEQ 
\rho_B&=&
\frac{1}{8\pi}\left(\frac{B''}{V}+\frac{2rB'}{W^2}-\frac{B'U'}{2UV}\right)
,\nn\\
p^B_r&=&\frac{-1}{8\pi}\left(\frac{B'U'}{2UV}+\frac{2B'W'}{VW}\right), \\
p^B_\perp
&=&\frac{-1}{8\pi}\left(\frac{B''}{V}+\frac{2rB'}{W^2}-\frac{B'W'}{VW}\right).\nn
\EEQ
Eq. (\ref{EinBeqs}) now leads to a total energy momentum tensor
\BEQ \label{TmnBr}
T^\mn_{\rm tot}=\frac{T_m^\mn+T^\mn_\Lambda+T_\bi^\mn+T_B^\mn}{1+B}\equiv 
T^\mn+T^\mn_\Lambda+T^\mn_\bi.
\EEQ
 $T^\mu_\nu\equiv{\rm diag}(\rho,-p_r,-p_\perp,-p_\perp)$ has at $\lambda=0$ the elements
\BEQ 
\rho=\frac{1}{8\pi(1+B)}
(\frac{2B'W'}{VW}-\frac{B'V'}{2V^2}+\frac{B''}{V}+\frac{m^2B}{2\xi}
+\frac{B'{}^2}{8\xi BV}), \nn 
\EEQ\vspace{-4mm} \BEQ 
 \label{rhopB}
p_r=\frac{1}{8\pi(1+B)}(
-\frac{2B'W'}{VW}
-\frac{B'U'}{2UV} %-\frac{3B}{4M^2}
-\frac{m^2B}{2\xi}
+\frac{B'{}^2}{8\xi BV}),
\EEQ\vspace{-4mm}\BEQ
 p_\perp=\frac{1}{8\pi(1+B)}
(\frac{B'W'}{VW}-\frac{2rB'}{W^2}-\frac{B''}{V}
-\frac{m^2B}{2\xi}
-\frac{B'{}^2}{8\xi BV}).
\nn
\EEQ
The Gross-Pitaevskii equation (\ref{GPeq}) reads in terms of $B$
\BEQ \label{fullBeqn}
&& -[(6\xi+1) B+1](\frac{B''}{2V} +\frac{rB'}{W^2})
+\frac{B'{}^2}{4BV}+m^2B(1-B)
\nn\\&&+\lambdaB m^2 B^2=(E_0^2-\half\xi\mu^2)\frac{B}{U}.
\EEQ
We can now first verify that the total energy momentum
tensor is conserved due to the harmonic condition (\ref{harm}).
The singular term $B/U$ also drops out from (\ref{fullBeqn}) for 
\BEQ \label{E02=}
E_0^2=8\pi\xi\rho_\bi=\half\xi\mu^2.
\EEQ
This was already used in (\ref{rhopB}) to cancel the $1/U$ terms.
With $\xi\equiv 2m^2M_1^2/3$ and $\bar\mu\equiv\mu M_1$ it implies again $E_0=\mub m/\sqrt{3}$.
For very small $B\ll 1/\xi\sim 10^{-54}$ there will be an exponential fall off, 
$B\sim\exp(-\sqrt{5}mr)$, so the horizon is 
a fraction of the Compton length thick.
In this narrow range $\rho$, $p_r$ and $p_\perp$ vanish smoothly.
In the regime $B\gg1/\xi$, Eq. (\ref{fullBeqn}) simplifies and actually reduces 
to Eq. (\ref{GScond}),

\BEQ \label{eqnBshort}
-\frac{B''}{V}-\frac{2rB'}{W^2}=\frac{(1-\lambdaB )B-1}{2M_1^2}.
\EEQ
Here we can consider $B$ as vanishing sharply, $B\sim r_h-r$,
so that $T^\mn_B\neq 0$, keeping the ultimate exponential tail 
and decay of $\rho$ and the $p_i$ in mind.  Eq. (\ref{rhopB}) then brings 
\BEQ 
& \rho=&\frac{-1}{8\pi(1+B)}(\frac{B'U'}{2UV}-
\frac{4+2B+\bar\lambda(4B+3B^2)}{8M_1^2}),
\nn\\
\label{rhopB'}
& p_r=&\frac{-1}{8\pi(1+B)}(\frac{2B'W'}{VW}
+\frac{B'U'}{2UV}+\frac{6B+3\bar\lambda B^2}{8M_1^2}),\qquad\\
& p_\perp=&\frac{1}{8\pi(1+B)}
(\frac{B'W'}{VW}-
\frac{4+2B+\bar\lambda(4B+3B^2)}{8M_1^2}). \nn
\EEQ
Let us first return to the interior where $U=2\mub^2x^2$, $V=8M_1^2x'{}^2/y^2$
and $W=2M_1x$. Eq. (\ref{eqnBshort}) can be written as
\BEQ \label{Beqnx}
\frac{1}{4}x^2B_{yy}-yB_y+(1-\lambdaB )B=1.
\EEQ
To understand the structure of the problem,
we again take $\lambda=0$.
Then for any $A$ there is the solution
\BEQ \label{BA=}
B(x)=1+Ay=1+A\sqrt{1-x^2},
\EEQ
Expressing the shapes (\ref{rhopB'}) in $y$, we have 
\BEQ 
& \rho=- p_\perp=&\frac{1}{64\pi M_1^2(1+B)}(2B+yB_y+4),
\nn\\
\label{rhopBx}
& p_r=&
\frac{-3}{64\pi M_1^2(1+B)}(2B-yB_y).
\EEQ
Surprisingly,  their $A$-dependence factors out, keeping
a vacuum equation of state $\rho=- p=3/64\pi M_1^2$, 
so (\ref{BA=}) is a non-uniform, exact solution of the same metric. 
The horizon $B=0$ is now located at $r_h>r_1$ where $y_h=-1/A$.
[Eq. (\ref{rx=}) continues   to negative $y\approx\sqrt{5}(r_1-r)/4r_1$ for $r>r_1$].
However, a problem shows up with the matching, since $W'(r_h)\sim-1/A$ 
cannot be of order $\mub\sim 10^{-14}$  anymore. 
We thus have to deviate from the exact solution,
which leads in general to a numerical problem.
Analytically, this question can be considered for large $A$, by adding $1/A^2$ corrections
to previous solution. Expanding in $1/A$ at fixed $s\equiv A\sqrt{5}(r/r_1-1)/4$,  
we arrive at
\BEQ
B&=& 1-s+\frac{\sqrt{5}}{2A}s^2 -\frac{7s^3}{6A^2}+\frac{b_1(s)}{A^2},\quad 
\nn\\ 
\frac{U}{ 2\mub^2}&=&1-\frac{s^2}{A^2} +\frac{u_1(s)}{A^2}, 
\qquad
\nn\\ 
\frac{2}{5}V&=& 1-\frac{2\sqrt{5}}{A}s+\frac{13s^2}{A^2}+\frac{v_1(s)}{A^2},
\nn\\  \frac{W}{2M_1}&=& 1-\frac{s^2}{2A^2}+\frac{w_1(s)}{A^2}.
\label{WA=}
\EEQ
The additional terms, found to be 
\BEQ
\label{Acorrs}
&b_1= b_{10}+b_{11}s,\qquad 
&u_1=u_{10}+4w_{11}\ln(2- s), \nn\\
&v_1= v_{10}, \qquad \qquad\quad
&w_1=w_{10}-w_{11}\ln(2- s), \qquad
\EEQ
produce an anisotropy, $\rho\neq -p_r\neq -p_\perp\neq\rho$.
The horizon $B=0$ is located at $s_h=1+\sqrt{5}/2A+{\cal O}(1/A^2)$, where
$W'=\sqrt{5}(w_{11}-1)/2A$.
Clearly, $W'\sim\mub$ from (\ref{vacuumvalues}) can be attained 
by tuning $w_{11}=1+3A\mub+{\cal }O(1/A)$.
The maximum of $W$ at $s=0$ in the absence of the $w_1$ term has 
now been shifted to the horizon, which shows that the problem has a proper solution, 
with $A$ remaining a free parameter.  $U$ has a maximum at $s=1-\sqrt{3}$. 
The mass seen at infinity, $M\equiv\M(r_h)=\half W(r_h)$, coincides with $M_1$ to order $1/A^2$.
At the horizon we can fix $r_1$ from (\ref{scaling}),
\BEQ 
r_1=M_1\left[1-\frac{4}{\sqrt{5}\,A}+\frac{4\mub}{\sqrt{5}}(2+\ln2\mub^2)
+{\cal O}(\frac{1}{A^2})\right].
\EEQ  

\subsection{Properties of the solution}
The groundstate occupation number becomes upon neglecting the $1/A^2$ corrections

\BEQ N_0=\int \d{\cal V} |\Psi_0|^2=2\sqrt{3}\,\frac{M}{m}\int_{-1/A}^1\d y\,(1+Ay).
\EEQ
We may write the two leading orders as
\BEQ M=\nu N_0m,\qquad \nu=\frac{1}{\sqrt{3}(2+A)}
\approx \frac{1}{\sqrt{3} \,A}. \EEQ
Clearly, the energy $Mc^2$ of the BH can be any fraction of the rest energy 
$N_0mc^2$ of the constituent hydrogen atoms
If $\nu$ starts at a value $\nu_c<1$, our BH is likely approached in an explosive manner,
possibly related to jets of quasars.

We found the sharpness of the horizon to be a fraction of the Compton length of H.
On a much larger scale $\ell_{\rm grav}$ there is  
near-horizon growth of the metric functions $U/\eta\approx \eta V\approx e^\xi$, 
taking place in the millimeter range,
\BEQ 
\ell_{\rm grav}=\frac{\d r}{\d\xi}=2\eta M= \frac{4}{\sqrt{5}}\,\mub\, M
=  4.88\,10^{-3} \,\frac{M^2}{M_\star^2}\,{\rm m}.
\EEQ 
It is a realistic value, small compared to the size of the BH, 
and still large compared to the Bohr radius.

\subsection{The interacting situation} 
If the nonlinearity $\lambda$ is relevant, 
a numerical solution is called for. Assuming the same leading order 
behaviors near $r=0$, the above structure survives.
Mass and particle number remain independent parameters.
This analysis remains as a task for future.

\subsection{Conclusion} 
We have questioned the general wisdom that static BHs have all their
mass in the center and cannot be described by present theories.
Numerical estimates show that a picture of closely packed H
atoms naturally applies to the supermassive BH's in the center of galaxies, 
$M\sim M_\star=2.34\, 10^8M_\odot$.
We present within the Relativistic Theory of Gravitation 
 (RTG; a colloquial term is: Not-so-General Relativity),
an exact solution for a BH, of which the interior is governed
by quantum matter in its Bose-Einstein condensed phase.
Its density decays algebraically in the bulk and exponentially near the horizon.
This solution is matched with the Schwarzschild metric, which 
near the horizon is deformed in RTG. 
Powerlaw singularities occur at the origin, that get absorbed in the Riemann description 
of the metric. Elsewhere, the solution is regular. The redshift at the horizon is
finite, though of the order $1/\mub \sim 10^{14}M_\star/M$.
To specify a BH requires not only the mass (and, in general, charge and spin),
but also the rest energy of the constituent matter.

Our BH is a quantum fluid confined by its own gravitation.
In the interior, time keeps its standard role. 
No Planckian physics is involved; Hawking radiation is absent 
and Bekenstein-Hawking entropy plays no role. 

Our BH has one ``hair''.
As one would expect for a classical theory of gravitation, when the quantum matter in the
BH has reached a certain groundstate, the classical metric allows the system 
still to go to a lower energy state.
Indeed, the passage of celestial bodies will induce oscillations in the metric
and emission of gravitational waves, which, upon re-equilibration, increase the binding energy, 
finally up to $100\%$ of the rest energy of its constituents, $Nmc^2$.
This property may explain the enormous jets and energy output of quasars and also be responsible 
for very high energy cosmic rays, $E> 4\,10^{19}$eV.~\cite{Auger}

An important question is whether formation of realistic supermassive BHs 
brings the matter indeed in or near the Bose-Einstein condensed groundstate. 
Also the stability of the solution needs to be studied.
It also remains to be seen whether the phenomenological value for the parameter $\xi$
has a microscopic underpinning.
Extension to finite temperatures, not presented here, 
will exhibit a $T^{3/2}$ fraction of thermal atoms. 

Calculation of the normal mode spectrum may lead to predictions that deviate from 
the ones of GTR; this spectrum may be observed in the foreseeable future.

If we apply Eq. (\ref{E02=}) to Logunov's case $\rho_\bi<0$, it follows that $\xi<0$, 
so $B<0$ due to (\ref{Bdef}). To avoid a singularity in e.g. (\ref{rhopB}), 
a lower bound $B(0)>-1$ will be required. 
As already indicated by the exact solution (\ref{BA=}),
$B$ will then go to zero at some point well below $M$. This prevents
fitting to the external metric and excludes our BH solution.

We failed to apply our approach to GTR, technically because it lacks 
compensation for the singular $1/U$ terms. If no other solution exists for the 
considered physical situation, GTR must be abandoned and replaced by another theory, RTG
being the first candidate. In view of its smaller symmetry group, this may have far 
reaching consequences for singularities in classical gravitation and 
for quantum approaches to gravitation, while Minkowski space-time needs no quantization. 

\acknowledgements  The author has benefited from discussion 
with Steve Carlip, Kostas Skenderis and Bahar Mehmani.

\end{document}